%% file: paper.tex
\documentclass [10pt] {revtex4}
\usepackage {amsmath}
\usepackage {amstext}
\usepackage {latexsym}
\usepackage {xspace}

\begin{document}


\newcommand\DateYMD{
	\renewcommand\today{\number\year.\number\month.\number\day}
	}

\newcommand\DateDMY{
	\renewcommand\today{\number\day.\number\month.\number\year}
	}

\newcommand\DateDmmmY{
	\renewcommand\today{
		\number\day\space
		\ifcase\month\or
		Jan\or Feb\or Mar\or Apr\or May\or Jun\or
		Jul\or Aug\or Sep\or Oct\or Nov\or Dec\fi
		\number\year
		}
	}

\newenvironment{mptbl}{\begin{center}}{\end{center}}
\newenvironment{minipagetbl}[1]
	{\begin{center}\begin{minipage}{#1}
		\renewcommand{\footnoterule}{} \begin{mptbl}}%
	{\vspace{-.1in} \end{mptbl} \end{minipage} \end{center}}

\newif\iffigavailable
	\def\figavailable{\figavailabletrue}
	\def\nofigavailable{\figavailablefalse}
	\figavailable

\newcommand{\Fig}[4][t]{
	\begin {figure} [#1]
		\centering\leavevmode
		\iffigavailable\epsfbox {\figdir /#2.eps}\fi
		\caption {{#3}}
		\label {f:#4}
	\end {figure}
}

\newlength{\defitemindent} \setlength{\defitemindent}{.25in}
\newcommand{\deflabel}[1]{\hspace{\defitemindent}\bf #1\hfill}
\newenvironment{deflist}[1]%
	{\begin{list}{}
		{\itemsep=10pt \parsep=5pt \topsep=0pt \parskip=10pt
		\settowidth{\labelwidth}{\hspace{\defitemindent}\bf #1}%
		\setlength{\leftmargin}{\labelwidth}%
		\addtolength{\leftmargin}{\labelsep}%
		\renewcommand{\makelabel}{\deflabel}}}%
	{\end{list}}%

\makeatletter
	\newcommand{\numbereqbysec}{
		\@addtoreset{equation}{section}
		\def\theequation{\thesection.\arabic{equation}}
		}
\makeatother


\DateYMD			
\def\figdir{_figs}		



\def\arcdeg{\hbox{$^\circ$}}
\newcommand{\bold}[1]{\mathbf{#1}}
\newcommand\degree{$^\circ$}
\newcommand{\Qed}{$\bold{\Box}$}
\newcommand{\order}[1]{\times 10^{#1}}
\newcommand{\Label}[1]{\ \\ \textbf{#1}}
\newcommand{\Prob}[1]{\mathrm{\mathbf{Pr}}[#1]}
\newcommand{\Expect}[1]{\mathrm{\mathbf{E}}[#1]}


\newcommand\ala{\emph{a la}\xspace}
\newcommand\vs{\emph{vs.}\xspace}
\newcommand\enroute{\emph{en route}\xspace}
\newcommand\insitu{\emph{in situ}\xspace}
\newcommand\viceversa{\emph{vice versa}\xspace}
\newcommand\terrafirma{\emph{terra firma}\xspace}
\newcommand\perse{\emph{per se}\xspace}
\newcommand\adhoc{\emph{ad hoc}\xspace}
\newcommand\defacto{\emph{de facto}\xspace}
\newcommand\apriori{\emph{a priori}\xspace}
\newcommand\Apriori{\emph{A priori}\xspace}
\newcommand\aposteriori{\emph{a posteriori}\xspace}
\newcommand\nonsequitor{\emph{non sequitor}\xspace}
\newcommand\visavis{\emph{vis a vis}\xspace}
\newcommand\primafacie{\emph{prima facie}\xspace}

\newcommand\circa{\emph{c.}\xspace}
\newcommand\ibid{\emph{ibid.}\xspace}		
\newcommand\loccit{\emph{loc.\ cit.}\xspace}	
\newcommand\opcit{\emph{op.\ cit.}\xspace}	
\newcommand\viz{viz{}\xspace}			

\newcommand\ie{i.e.{}\xspace}			
\newcommand\eg{e.g.{}\xspace}			
\newcommand\etal{\emph{et al.}\xspace}		
\newcommand\cf{cf.{}\xspace}			
\newcommand\etc{etc.{}\xspace}			

\newcommand{\figref}[1]{Fig.\ (\ref{f:#1})}
\newcommand{\figsref}[2]{Figs.\ (\ref{f:#1}-\ref{f:#2})}

\newcommand{\LD}{\begin{description}}
\newcommand{\DE}{\end{description}}
\newcommand{\LI}{\begin{itemize}}
\newcommand{\LE}{\end{itemize}}
\newcommand{\LN}{\begin{enumerate}}
\newcommand{\NE}{\end{enumerate}}
\newcommand{\VB}{\begin{verbatim}}
\newcommand{\VE}{\end{verbatim}\\}
\newcommand{\QB}{\begin {quotation}}
\newcommand{\QE}{\end {quotation}}

\newcommand{\Or}{\vee}
\newcommand{\Def}{\stackrel{\triangle}{=}}

\def\ednote#1{\noindent== \emph{#1} ==}		
\def\note#1{\noindent====\\\emph{#1}\\====}	
\def\revising{\ednote{to be revised}}
\def\comment#1{}			


\title {
	Relativity of spatial scale and of the Hubble flow:\\
	The logical foundations of relativity and cosmology
}
\author {V Guruprasad}
\affiliation{IBM T J Watson Research Center,Yorktown Heights, NY 10598}
\email {prasad@watson.ibm.com}
\begin {abstract}
Formalising
the \emph{logical} dependence of physical quantities
on material referents of scale,
I show that
both Hubble's law and the cosmological constant
are in fact \emph{exactly} replicable by
a spatial contraction of referents locally on earth, and that
the Pioneer anomaly is irrefutable indication
that this is the case.
The formalism literally embodies Feynman's ``hot-plate'' model,
predictably yielding
a logical derivation of the relativity postulates,
and importantly, suffices to demonstrate
the inherent logical consistency of
general relativity and quantum mechanics,
whose foundations I have shown separately
to be fundamentally computational.
I further show that
the spatial contraction of our referents also accounts for
considerable planetary and geological data
inexplicable in the standard model, and
predicts that the Hubble flow appears differently or
is altogether absent from platforms in deep space,
depending on the local physics.
\end {abstract}
\maketitle
\newcommand\Prime[1]{#1{}'}
\newcommand\s{s$^{-1}$}
\newcommand\ssq{s$^{-2}$}

\section {Introduction}
\label {s:intro}

Traditional reasoning holds that
a simultaneous expansion of
the observer's own body and instruments
would render the cosmological expansion unobservable
\cite [p719] {MTW}
\cite [p197] {Rindler}, and that accordingly,
the observable expansion cannot include small scales.
It has been lately shown that
while the expansion could occur on all scales
\cite {Anderson1995},
the unobservability argument does not get challenged,
as even over planetary distances,
the effects would be extremely small
\cite {Cooperstock1998}.

I examine here the converse question:
shouldn't, then, a \emph{contraction} of the observer,
not driven by the metric, be considered at least
as a mathematically equivalent picture of the Hubble flow?
I show that
this would not only reproduce the Hubble redshift in a static universe,
but would \emph{exactly} predict both
the cosmological constant $\Lambda$ (\S\ref{s:earth}) and
the Pioneer anomaly (\S\ref{s:anom}),
also obviating, in the first case,
the current speculations of a large-scale repulsion.
The hypothesised contraction also turns out to be consistent with
unresolved evidence of a past expansion of the earth
\cite {Wesson1973} (\S\ref{s:earth}),
which is irreconcilable with any of the past theories.

In view of the erroneous objections raised by various referees,
I need to point out that
the prior relativistic notion of scale,
even in the Brans-Dicke theories,
is by definition defined by the relativistic metric, and
signifies only a general order of magnitude of the observed entities
\cite [p98] {Wald},
\ie a \emph{virtual} scale,
in which the observer's own neighbourhood,
including its physical unit referents of scale, remains unscaled.
This makes the existing framework of general relativity
\emph{logically incomplete},
as there is no way to account for mechanisms
that might cause our unit referents to vary differently
from the underlying metric, and
the absence of such mechanisms cannot be considered \apriori
to be a law of physics.
I have separately described
the complete logical foundations of quantum mechanics
\cite {Prasad2000b}, and
shown that the constancy of Planck's constant $h$
does not suffice to establish equality of spatial scale
from either quantum interactions or thermal equilibrium.
Formalising the dependence on referents not only
leads to a more precise interpretation of
relativistic curvature (\S\ref{s:scale}),
but also to a fundamental and appropriately simple proof of 
the inherent consistency with quantum mechanics
(\S\ref{s:scale}, Appendix \ref{a:rel}), and
to a purely logical derivation of the relativity postulates.

A second clarification is also clearly in order,
concerning the obvious difficulty of accounting for
a \emph{source}-dependent redshift
purely by the \emph{observer}'s physics, which is why
cosmological expansion and ``tired''-light theories
have been the only explanations conceived of.
These theories are, however, incomplete for the same reason,
\viz that variability of the spatial scale of
the observer's instruments was not recognised in
the prior formulation of quantum mechanics,
which had evolved by
matching conjectural notions with empirical grounds,
rather than of sound reasoning from
the fundamental definitions of mechanics.
The issue is therefore unaddressed by relativistic field theory,
which accounts only for the relativistic metric, and
not possible non-metrical variation of referents.
As particularly established now in
the classical thermodynamic derivation of Planck's law
\cite {Prasad2000a},
the radiation quantum corresponds to an incremental change of
an antinodal lobe in a stationary mode of the receiver, and
the energy of a lobe \perse is independent of its wavelength,
meaning that
quantisation does not require the equality of spatial scale
between interacting entities.
The packetisation of incoming electromagnetic energy thus depends
\emph{entirely} on the thermodynamics of the observer.

\section {Relativity of Hubble's law}
\label {s:form}

It is remarkable that
despite the unobservability argument cited above and
Feynman's ``hot-plate'' model
suggesting the variability of material referents 
\cite [II-42-1] {Feynman},
no formal consideration of the implied dependence on local referents
exists in the founding concepts of relativity
\cite [ch.1] {EinsteinMeaning},
or in the subsequent literature.
The dependence is definitional,
because even to conceive of a physical quantity $S$,
unless it be a dimensionless ratio,
we need reference to some material referent $R$.
For example,
in citing the wavelength of a Lyman-$\alpha$ line,
we implicitly refer to the standard metre,
itself defined as so many wavelengths of an atomic transition.
In attributing the redshifts of stellar Lyman spectra
to Doppler and gravitational effects only,
we implicitly assume that
terrestrial and stellar matter are otherwise inherently identical.
Although this looks entirely reasonable,
our physics can be neither logically complete nor precise
without formally accounting for this circularity and
examining how it could break our assumption. 
This formalisation is readily obtained
from the simple observation that
the numerical value of a quantity $S$ is necessarily a ratio
\begin {equation} \label {e:scale}
	n = f (S) / f (R) ,
\end {equation}
where
$f$ denotes the ``notch-counts'' obtained
in the physical course of measurement, and
$f(R)$ signifies the calibration.

The possibility of virtual Hubble flow,
which does break our past assumption, now arises as follows.
The traditional derivation of Hubble's law
\cite [p98] {Wald},
\begin {equation} \label {e:hubble}
	v (S)
	= \frac{dS}{dt}
	\equiv
		\frac{S}{a_v}\frac{da_v}{dt}
	= H \, S ,
		\quad
		H \equiv \frac{\dot{a}_v}{a_v} ,
\end {equation}
concerns a ``virtual'' form of scale
not relating to material referents $R$ or changes thereof
(hence my subscript $v$),
\ie $R$ has been implicitly assumed to be constant in the past.
Recalling from Dirac's Large Number Hypothesis (LNH)
\cite {Dirac1937}
that even the constancy of the absolute constants
cannot be assumed on this time scale,
there can be justification for assuming $R$ to be constant
on this scale either.
Incorporating eq.\ (\ref{e:scale}) into eq.~(\ref{e:hubble}),
and allowing $R$ to vary, we get
\begin {equation} \label {e:deriv}
\begin {split}
	n(v (S))
	&\equiv
		\frac{d}{dt} \frac{f(S)}{f(R)}
	=
		\frac{1}{f(R)} \frac{df(S)}{dt}
		- \frac{f(S)}{f(R)^2} \frac{df(R)}{dt}
\\
	&=
		\frac{1}{f(R)} \frac{df(S)}{dt}
		- \frac{f(S)}{f(R)}
			\left[
			\frac{1}{f(R)} \frac{df(R)}{dt}
			\right]
	,
\end {split}
\end {equation}
more succinctly expressible as
\begin {equation} \label {e:chubble}
\begin {split}
	v (S)
	= \frac{dS}{dt} - S \frac{da}{dt} 
	= \frac{S}{a_v} \frac{da_v}{dt} - \frac{S}{a} \frac{da}{dt}
	&= H_t \, S ,
\\
\text{where }
	H_t
	&= H - H_r ,
\text { and }
	H_r
	\equiv \dot{a} / a
\end {split}
\end {equation}
for the observed (total) redshift $H_t$,
where $H$ continues to represent actual expansion;
the unsubscripted $a$,
a scale factor expressly referencing our ``real'' referent $R$; and
$H_r$ quantifies the impact of $R$'s expansion on the observation.
This is a clear and correct formalisation of the relativistic argument,
as by setting $\dot{a} / a = \dot{a}_v / a_v$,
$a_v$ denoting the scale factor in the 
the Friedmann-Robertson-Walker (FRW) formalism,
we do get $H_t = 0$ as expected.

Eq.\ (\ref{e:chubble}) also admits, however, the possibility that
\emph{any local mechanism capable of causing
a continuous variation of $R$ ($\sim \dot{a} \ne 0$)
could introduce redshifts obeying Hubble's law}.
For example,
the atomic wavelengths used to define the metre
\cite {Petley}
would necessarily change due to incremental gravitational redshift
if the earth's mean radius $r_e$ were to be changing for some reason,
which has been hypothesised before
\cite {NarliKem1988}
\cite {Wesson1973}.
By the circularity described above,
the change would be undetectible on ground, as well as by satellites,
as their orbital heights are dependent on the same ground referents.
The circularity is broken only by the deep space missions,
which is why the deep space anomaly acquires a fundamental significance
to be described in \S\ref{s:anom}.

\section {Gravitational shrinkage}
\label {s:scale}

As the constancy of $R$ is a deeply ingrained belief
\cite [p3] {Dirac},
I need to first show that
such a variation is produced by relativity
as a result of gravitational redshift.
The basic idea is that if
the internal scale $a$ were indeed independent of
the electromagnetic wavelengths,
the structure of matter should become incongruous with radiation
in the presence of gravity.
This is particularly suggested by
the Huygens picture of the gravitational bending of light;
specifically,
the redshift is traditionally interpreted as scale-preserving,
but the notion depends on assuming that $c$ varies
\cite {Einstein1911}.
To appreciate this and the present result,
recall that in special relativity,
a light clock essentially relates
the local scales of length and time, and
that there is no \apriori reason for assuming
either to remain absolutely constant.
We could take an equivalent view
in which $c$ is preserved and the length scale varies instead,
as follows.

Consider two observers bearing light clocks,
the first, subscripted $1$, stationed within a gravitational well and
the other, subscribted $2$, located outside,
their light clocks being of lengths $l_i$ and periods $t_i$,
$i = 1, 2$, respectively.
In terms of our more precise formalism of eq.~(\ref{e:scale}),
special relativity postulates that $c$ is locally preserved, \ie
\begin {equation} \label {e:cequality}
	\frac{ f_1 (l_1) }{ f_1 (t_1) }
	=
	\frac{ f_2 (l_2) }{ f_2 (t_2) }
	= c ,
\end {equation}
where the $f_i$ denote notch-counts by the $i$-th observer.
In Einstein's variable-$c$ perspective,
clock $2$ runs slower than clock $1$,
meaning $f_i (t_2) > f_i (t_1)$.
This merely implies
$f_2 (l_2) / f_2 (t_2) <  f_2 (l_2) / f_2 (t_1)$
and is insufficient to establish
\begin {equation}
		\frac{ f_2 (l_2) }{ f_2 (t_2) }
	\equiv
		c
	<
		c_{1,2}
	\equiv
		\frac{ f_2 (l_1) }{ f_2 (t_1) }
	,
\end {equation}
$c_{i,j}$ denoting the speed of light
at $i$-th location measured by observer $j$,
for which we also need to assume $f_2 (l_2) = f_2 (l_1)$.
There is no physical basis for this further assumption:
the only \emph{physical} relation we have is already given by
eq.\ (\ref{e:cequality}),
with which all physical laws are presumably consistent,
so $c_{i,j}$ cannot bear independent significance to physics.
It should be entirely equivalent, therefore, for us to also take
\begin {equation} \label {e:lineq}
	c_{i, j} = c
\text { so that }
	f_i (l_2) > f_i (l_1) .
\end {equation} 

This is a remarkable result,
as illustrated by application to Schwartzschild geometry.
Say our unit rod is a solid $N$ lattice constants in length;
by eq.\ (\ref{e:lineq}),
its atoms should become more densely packed
when the rod is taken into the earth.
Direct measurement of its diameter by lining up unit rods
should thus yield a larger number than
that obtainable from the surface area or circumference,
which would be measured by rods left on the surface.
A simpler explanation of gravitational curvature is thus achieved,
which is not only logically more precise than prior theory,
but particularly brings out
the conservation of quantised properties like the number density $N$.
The result represents \emph{covariance} of the quantum scale, and
is clearly a less restrictive interpretation than constancy and
the only one required by the laws of physics.
Further support for
this inherent quantum consistency is given in Appendix \ref{a:rel},
where the very postulates of relativity
are logically derived from these notions.

The reality of the \emph{virtual} Hubble flow, $H_r$,
is also easily established by this equivalence.
If observer $1$ is subject to an increasing acceleration $\dot{g}$,
eq.\ (\ref{e:lineq}) says that
its local rods and clocks must shrink at rate
$H_r \equiv \dot{a}/a < 0$;
all objects will then appear to be receding at
velocities proportional to their distance,
\emph{replete with Doppler shift resembling the cosmological expansion}.
This conflicts with our usual notions,
since, from the perspective of observer $2$,
identical photons from different sources,
bearing no information of the respective distances,
appear to be magically received by observer $1$ at varying redshifts
\cite {Elmegreen1999pvt}.
It is, however, as should be, because
the absorbed photon eigenfunctions must correspond to
the stationary states of the receiver in
the receiver's reference frame, and
those of observer $1$ cannot be stationary with respect to $2$.
As observer $1$'s stationarity involves a decreasing $c$,
relative to $2$,
its atoms cannot ``know'' of this decrease,
let alone that it is a local artifact, and
must therefore perceive the optical path lengths
as inexplicably increasing.
The applicable eigenfunctions have long been described by Parker
in another context
\cite {Parker1968}
\cite {Parker1969}
and do display the distance-dependence.
\Qed

\section {Variable earth implications}
\label {s:earth}

A changing gravitation $\dot{g}$ on earth,
for instance due to changing $G$ or $r_e$,
could thus contribute to the Hubble redshift.
Certain geological and paleomagnetic evidence have indicated
a past expansion of the earth at about $0.4$-$0.6$~mm/y
\cite {Runcorn1965},
which is still about two orders of magnitude
too small to be directly verified.
Incidentally,
the indicated expansion is difficult to reconcile
with known physics, and
even variable-$G$ theories can account for
only a fraction of this rate
\cite {Wesson1973}.
It is also of the wrong sign to account for the Hubble redshift,
but fortunately too small to be significant:
from the incremental redshift, we find
$- \dot{a} t = \delta \Phi/c^2 \approx g r_e / c^2$,
so that the full Hubble flow corresponds to a contraction rate of
$-dr_e/dt
\approx c^2 \dot{a}/ g
\equiv c^2 H r_e / g
\approx 128.43$ km/s,
ruling out significant variable-$G$ and $\dot{r}_e$ contributions.
Moreover,
the circularity of scale prohibits
a \emph{uniform} expansion of all matter on earth
from being observable.

It does not, however, rule out
an ongoing contraction of \emph{surface matter},
which would include our solid referents,
due to a non-relativistic cause,
as that could be just right, via eq.\ (\ref{e:scale}),
to cause the earth to \emph{appear}
to have expanded at the empirically indicated rate.
The possibility is valid as the evidence equivalently indicates that
the sialic masses contracted \emph{relative} to $r_e$
\cite {Wesson1973}, but
it also implies a proportional recession of the moon and the planets,
as well as of the distant stars.
The past expansion indeed happens to be of the same order as $H$
\cite {MacDougall1963}
($0.4~\text{mm/y} \approx 65~\text{km/s-Mpc}$), and
even the lunar recession and existing planetary range data
are consistent with the small-scale Hubble flow,
(see Appendix \ref{a:planetary}).
Additionally,
an upper bound of $z = 2$ exists for gravitational redshift
\cite [\S6.3] {Wald},
which, by our analysis in \S\ref{s:scale},
would impose an upper bound on the redshift of the incoming photons;
this too is clearly represented in the geological data,
as $r_e$ appears to have at most doubled
since the birth of the solar system
\cite {Runcorn1965}.
As mentioned,
the contraction also exactly predicts $\Lambda$, as follows.

Hubble's law $\dot{r} = H_t r$
implies an intrinsic acceleration as
a particle initially at distance $r$
must pick up the speed increment
$\delta \dot{r} \equiv H_t \, \delta r$
by the time it reaches $r + \delta r$.
This intrinsic acceleration
$\ddot{r} = \dot{H}_t r + H_t \dot{r} \equiv \dot{H}_t r + H_t^2 r$
already represents the deceleration factor
\begin {equation} \label {e:qfactor}
	q \equiv - \frac{\ddot{a}}{a} \frac{1}{H_t^2}
	= - \frac{1 + \dot H_t/H_t^2}{a}
	= - (1 + \dot H_t/H_t^2) .
\end {equation} 
Now,
$\dot{H}_t \equiv \dot{H} - \dot{H}_r$, and per FRW theory,
$\dot{H}$ has many possible variations depending on
the undetermined curvature constant $k$ and
the radiation pressure
\cite [p98] {Wald}
\cite [p772-774] {MTW},
allowing for the well known gamut of values for
the matter ($\Omega_M$) and energy ($\Omega_\Lambda$) densities
in the universe
\cite {Reiss1998}
\cite {Garnavich1998a}.
However,
eq.\ (\ref{e:chubble}) and its underlying semantics (eq.\ \ref{e:scale})
bear no dependency on the past values of $H_r$, meaning that
$H_r$ would always appear to be constant at the instant of measurement,
\ie $\dot{H}_r = 0$ regardless of the past values of $H_r$.
Thus,
if the observed Hubble flow $H_t$ were indeed due to $H_r$ alone,
we would get $q = -1$ identically,
which is precisely the value
consistently indicated by Type Ia supernovae
\cite {Reiss1998}
\cite {Leibundgut1998}
\cite {Garnavich1998b}.
\Qed

\section {Deep space confirmation}
\label {s:anom} 

As mentioned in \S\ref{s:form},
measurements from deep space would be independent of
the inherent circularity of scale, and
should therefore be able to test the theory. 
Because of the immensely small order involved
($\sim H \equiv 10^{-18}$~s),
only the six deep space missions that employed spin-stablisation 
could provide ranging data of the necessary precision
\cite {Anderson1998}, and
\emph{all six} have displayed an anomaly of precisely this order
\cite {Anderson1998}.
Only a constant part, $h_c \approx H$,
is actually to be accounted for by $H_r$,
as the variations between missions and along the trajectories
are conceivably due to mundane mechanisms,
given that the temporal variations of $\dot{g} \propto - \dot{r}/r^2$
cannot account for its apparently oscillatory character
\cite {Turyshev1999}.
The inferred acceleration
$\delta g = r^{-1} \, c^2 \, \delta z$ is
stated to be quantitatively equivalent to the time dilation
\cite {Anderson1998}
\begin {equation} \label {e:hmeasure}
	h_c = c^{-1} \, \delta g
		\approx 2.8 \times 10^{-18}
			\text{ s}^{-1}
		\equiv 86
			\text { km/s-Mpc}
\end {equation}
signifying an expansion of onboard clocks
\emph{relative} to those on earth.
Unlike the relativistic $H$,
which is negligible at planetary range
\cite {Cooperstock1998},
$H_r$ should produce a cosmological time dilation (CTD)
as a uniform dilation of \emph{all} clocks in the universe
\emph{relative} to our own,
implying that the anomaly is indeed the CTD expected from $H_r$.
This has two fundamental implications,
first, that onboard the spacecraft,
the observable Hubble flow must be
$H_t \equiv H - h_c \approx 0$,
\ie
\emph{the Hubble flow must be invisible from deep space}; and
second,
that the relativistic $H$, from FRW theory, must be zero
as $H_t$ is almost the same for the distant stars,
\ie
\emph{the Hubble flow must be entirely due to terrestrial contraction.}

As a check, we may again write Hubble's law as $\dot{r} = h r$,
predicting a continually increasing unmodelled contribution
in the signal path $r \sim c t$ to the spacecraft,
with rate of change once again resembling an acceleration
\cite {Rosales1998}
\begin {equation} \label {e:accel}
	\delta g
	\equiv \ddot{r}
	= \frac{d}{dt} \dot{r}
	= \frac{d}{dt} \{ h c t \} = h c ,
\end {equation}
identical to NASA's computation of
the equivalent time dilation, eq.\ (\ref{e:hmeasure}).
Eq.\ (\ref{e:accel}) does not suffice to prove planetary Hubble flow,
as we cannot deduce $\dot{r} = h r$ from it.
But the consistency of $h_c$ across all six missions
that were at all equipped to measure it
\cite {Anderson1998},
our inability to explain the constant residual part
\cite {Turyshev1999}
\cite {Anderson1999a},
its consistency with lunar and planetary range data
(see Appendix \ref{a:planetary}) and
$\Lambda$
(\S\ref{s:earth}), and
the \emph{total} absence of evidence to the contrary,
together seem to be compelling indication
that this is the case.

Eq.\ (\ref{e:accel}) is not the complete picture
because unlike ordinary radar, the ranging procedure
also involves a significant onboard segment in the signal path,
comprising frequency conversion and phase-locked loops
\cite {Bender1989}
\cite {Vincent1990}
\cite {Anderson1993}
\cite {Anderson2000}.
The effective length of this segment cannot possibly be constant
to within the magnitude of the anomaly,
\viz $O(10^{-18})$~\s,
as plastic flow under centrifugal action alone
could cause expansion of at least this order, and
its modulation due to the varying orbit conditions
could account for the reported variations;
in particular,
this seems to explain the perihelion increase 
($\delta g \approx 12 \times 10^{-10}$~m/s$^{2}$ at $1.3$~AU,
in the case of Ulysses).
The mechanism should likewise produce contraction
under the compressive stress of earth's gravity, and
the tidal factors seem to correctly scale as well,
as will be described separately,
in strong support of the present theory.

\section* {Conclusion}
\label {s:concl}

As mentioned,
Feynman's ``hot-plate'' model anticipates the formalism  
introduced here.
Its power, demonstrated by
the logical derivation of the relativity postulates and
proof of the inherent consistency with quantum mechanics
(Appendix \ref{a:rel}),
reflects the improved logical precision in
the treatment of physical variables.
It should also be clear that
the cosmological constant $\Lambda$ and the Pioneer anomaly 
constitute two null indications, \viz
$\dot{H}_r = 0$ and $H_t = 0$ in space, respectively,
in analogy to the Michelson-Morley result,
that favour the present theory.

Admittedly,
we do not as yet know how or if other successes of the standard model,
such as the cosmic microwave background (CMB) and Olbers' paradox,
would be accomodated in the present theory,
but the concern seems to be outweighed by
the geological and deep space data near at hand.
Some of these questions might already be answered in
the ``scale-expansion cosmology'' (SEC) theory
\cite {Masreliez1998},
as the latter's conjecture of homogeneous expansion of scale
is exactly provided by $H_r$.
The predicted invisibility of the Hubble flow in deep space
also remains to be verified.

\acknowledgements

Many thanks are owed to Bruce G Elmegreen, A Joseph Hoane and
Gyan Bhanot for substantial criticisms, help
and guidance leading up to this work.



\appendix
\newcommand\sq{$^{2}$}
\section {Evidence of planetary flow}
\label {a:planetary}

I now show that the inference $H_t \approx -H_r \equiv H \approx 0$
is consistent with planetary, lunar and terrestrial data.
Anderson \etal contend that
the anomaly is missing in planetary ranging data,
but they have sought only
an actual acceleration in the planetary orbits,
using the Viking data to set the upper bound of 
$10^{-11}$~m/s\sq
\cite {Anderson1998}.
The absence of orbital influences invalidates
the acceleration theory,
but not the Hubble's law prediction given in \S\ref{s:anom}.
The observability of $H$ on planetary and shorter scales
has already been ruled out
\cite {Cooperstock1998};
what we need to examine is that of $H_r$,
which is non-relativistic and larger than $H$ by over $30$ orders.
Using eq.\ (\ref{e:chubble}), in place of the FRW equations,
we get recession rates of only
$1.6$~$\mu$m/s for Jupiter ($5$~AU) and
$12.5$~$\mu$m/s for Neptune or Pluto ($40$~AU),
far too small for Doppler measurement.
Even the yearly recession of Venus or Mars with respect to earth
would be only about $2$-$6$~m,
less than a tenth of the precision $100$-$150$~m cited by Anderson.
We would have to have measurements
spanning at least a $100$ years
to notice the cumulative effect at all.
Any ranging technique would also be inherently characterised by
a relative precision $\delta r \sim h_{\mu} r$,
\ie
\emph{the ranging error also follows Hubble's law}.
The anomaly was detected, therefore, only because
increased sensitivity by two orders of magnitude
intended to test general relativity
\cite {Bender1989}
\cite {Vincent1990}.

A more subtle problem is that
successive sets of measurement tend to be treated as improvements,
not comparisons,
so that cumulative displacements, if at all detected,
would have been attributed to ``systematic error'',
as described by Slade \etal for
the Goldstone radar observations of Mercury
\cite {Slade1998}.
A discrepancy of $+ 11$~km is in fact reported for Jupiter in 1992
from the predicted orbital radius by the PEP740 ephemeris
\cite {Harmon1994},
likely calibrated from 1970s data;
the discrepancy amounts to a recession of
$611$~m/y ($H \approx 770$~km/s-Mpc),
about $10$ times too large and enough to completely mask the recession.

Lunar and geophysical measurements are more precise and
just as much in favour.
We would have been contradicted by the lunar recession
if the latter were less than or almost equal to
our ``Hubble's law'' prediction ($2.57$~cm/y),
leaving no room to accomodate a purely frictional component.
However, the measured value is $3.84$~cm/y
\cite {Lunar1994},
almost $50$\% higher.
The slowing of the earth's rotation does not seem to be
an independent means for estimating the friction,
as it is already a factor in all the indications of
the past expansion of the earth
\cite {Runcorn1965}, and in any case,
the plastic flow mechanism (\S\ref{s:anom})
appears to be part of the tidal friction.
The indictated value $0.4$-$0.6$~mm/y
($\sim H \approx 61$-$92$~km/s-Mpc
\cite {MacDougall1963}),
is once again two orders smaller than
the available precision using the GPS,
which should, because of the circularity of scale,
be inherently incapable of detecting the expansion
(\S\ref{s:form}, \S\ref{s:earth}).
\Qed

\input {tbphub}

\section {Verification}
\label {a:veri}

It has been suggested
\cite {Elmegreen1999pvt}
that the theory could be verified on earth
by measuring the relative expansion, or ``aging'', of
a sufficiently strong laser pulse reflected back and forth
in an interferometer,
traversing a total optical path length $L$.
The resolving power of $\delta \phi$ fringes
would be matched within
\begin {equation} \label {eq:tFringe}
	\delta t = \frac{\lambda \, \delta \phi}{H_r L} = \frac {L}{c} ,
\quad \mathrm{or} \quad
	L = \sqrt{ \frac{c \, \lambda \, \delta \phi}{H_r} } .
\end {equation}
For example,
a shift of $10^{-6}$ fringes at $500$~nm requires $L = 7317$~km,
using the Pioneer anomaly as the indication of $H_r$.
With mirror reflectivity 99.9\%,
a 20~dB budget for mirror loss would allow 4600 reflections,
requiring the cavity to be $1.6$~km in length.
The phase shift would be distinguishable from
the measurement uncertainty in $L$ from
its $\lambda^{-1/2}$ dispersive character and
linear growth during the measurement, given by
\begin {equation}
	d \phi / dt = H_r L / \lambda .
\end {equation}
Any aging found should be entirely due to terrestrial contraction,
as FRW contribution should be zero on earth.
The method still presents several difficulties:
for instance,
the spectral spread of the pulse would also be amplified
by the same ratio, and
a refractive medium cannot be used to shorten the path,
as there would be no way to distinguish the effects of the medium
from true aging.
It might be possible to overcome some of these by using
matter waves instead of light,
as $\lambda$ could be made smaller by several tens of orders.

\newcommand\LT{\mathcal{L}}
\newcommand\GR{\mathcal{G}}
\newcommand\RT{\mathcal{R}}

\section {Logical deduction of relativity}
\label {a:rel}

I show below that
the formalism of \S\ref{s:form} also suffices for logically deducing
the postulates and both theories of relativity,
implying thereby that
the relativity of scale, eq.\ (\ref{e:scale}),
is the hitherto missing \emph{logical} basis
of the postulates of relativity.
Accordingly,
I deduce the postules solely on considerations of relative scale,
\viz
that observers separated by space or time cannot possibly
share their physical referents, and
that sharing is impossible 
even between coexisting, colocated observers
if they happen to be moving with respect to one another,
so that the only way to compare their observations
is by exchanging the numerical values from their measurements.
This leads to the Lorentz transform
by considering two such observers $O$ and $O'$,
as follows.

$O$ cannot assume that
$O'$ will arrive at the same numerical value
$n' \equiv f'(S)/f'(R')$
when measuring a space-time interval $S$,
as it cannot \apriori assume that
their ``notch-counts'' will match,
\ie in general, $f'(R') \ne f(R)$ and $f'(S) \ne f(S)$.
We thus have four unknowns, $R$, $R'$, $f$ and $f'$
for determining the relative scale, and
one common variable, the relative velocity $\bold{v}$,
whose magnitude must be match between their perspectives.
Each observer projects the other's measurements
on its own space-time plot, in particular,
obtaining a mapping $\LT$
of the other's coordinate axes onto its own.
As $\bold{v}$ is the only physical aspect distinguishing the observers,
$\LT$ must depend only on $\bold{v}$, and must be linear,
as a more complex transformation would require
additional distinguishing parameters, and
would imply different physical conditions
between their respective neighbourhoods.
The conditions of coexistence and colocation mean that
the origins can be trivially made to coincide,
\ie $\LT_{\bold{v}} (0,0) = (0,0)$, and
linearity means that the two sets of axes must be mutually inclined.
There are only two possibilities for this,
by a real or an imaginary rotation, and
the first is ruled out because it could
confuse the semantic distinction between space and time.
$\LT$ must be an imaginary rotation, therefore, and further, 
it cannot depend on $\bold{v}$ directly,
as we have not yet established the transformation of scales.
Accordingly,
we must have $\LT \sim \LT (\beta), \beta \sim \bold{v}/c$,
where $c$ has the same dimensions as $\bold{v}$.
We have thus deduced the \emph{form} of the Lorentz transformation,
and it remains for us to establish the constancy of $c$,
which presumably relates to $f$ and $f'$,
since the transformation must relate to
the cross-measurements of the referents, $f'(R)$ and $f(R')$.

To derive the special relativity postulates,
we first observe that
$\LT$ cannot be of fundamental significance
unless the physical interactions used in the measurement, $f$ and $f'$,
are themselves fundamental.
It would negate our purpose to \apriori assume
electromagnetism, or the strong or weak nuclear forces,
to be fundamental,
but such assumptions are unnecessary in the formalism,
as the relativity postulates follow logically 
from the notion of fundamentality:
\LN
\def\theenumi{\Alph{enumi}}

\item \label {i:creqn}
	The ``notch-counting functions'' must be \emph{analytic},
	\ie continuous in a path-independent manner,
	with respect to displacement of the \emph{observer}
	relative to the measured object $S$.
	As is well known from partial differential theory
\cite [\S62,65,66] {Sokolnikoff},
	analyticity over any two dimensions $x$ and $t$
	is expressed by the Cauchy-Riemann (CR) equations
\begin {equation} \label {e:CR2Div}
	\frac{\partial f_x}{\partial x} +
	\frac{1}{c}
	\frac{\partial f_t}{\partial t} =
	0
\text { and }
	\frac{\partial f_t}{\partial x} -
	\frac{1}{c}
	\frac{\partial f_x}{\partial t} =
	0 ,
\end {equation}
	where
	$f_x$ and $f_t$ are measuring functions along $x$ and $t$,
	the constant $c$ is simply
	a scale factor relating the dimensions.
	In our context,
	as $x$ is identifiable with space and $t$ with time,
	this $c$ has the requisite dimensions of speed and
	is identifiable with the Lorentz scale factor.
	Furthermore, the CR conditions yield the wave equation
\begin {equation} \label {e:CR2Wave}
	\frac{\partial ^2 \bold{f}}{\partial x^2}
-
	\frac{1}{c^2}
	\frac{\partial ^2 \bold{f}}{\partial t^2} = 0 ,
\text { where }
	\bold{f} = f_x , f_t
\end {equation}
	showing that
	the interactions $f$ and $f'$ could be used to communicate, and
	that the communication would be limited by the speed $c$.

\item \label {i:cmost}
	Given a multitude of such physical means
	that however differ in speed,
	we would eliminate all but the fastest interactions,
	as the information returned by a slow process
	would be rendered redundant
	if a faster process could be employed at the same time.
	As $f$ and $f'$ are physical interactions,
	this also applies to the transmission of physical effects.
	For example,
	sound would be eliminated because
	it is slower than electromagnetism, and
	in every case we can employ sound,
	electromagnetism is also involved
	in the form of intermolecular forces.
	Thus, $c$ must be the maximum of all physical speeds.

\item \label {i:csame}
	(\ref{i:cmost}) also implies that
	all fundamental physical means of interaction
	must have the same speed,
	as a fundamental interaction cannot be redundant.
	We would infer, for instance, that
	electromagnetic and gravitational waves
	must travel at the same speed $c$,
	if we knew they were both fundamental.

\item \label {i:cfinite}
	(\ref{i:cmost}) appears to favour an infinite value for $c$,
	but that would also mean, by (\ref{i:csame}),
	that all fundamental interactions
	would operate instantaneously.
	It would be impossible to construct
	any physical clock whatsoever,
	because \emph{no finite sequence} of
	fundamental physical interactions,
	meaning no realisable process,
	could yield a delayed action.
	An infinite speed of interaction is
	also functionally equivalent to colocation, hence
	\emph{for the dimensions of space and time to exist at all,
	the fundamental forces must operate at finite speed}.
	This proves $c < \infty$.

\item \label {i:cqm}
	Eqs.\ (\ref{e:CR2Div}-\ref{e:CR2Wave})
	clearly hold for quantum wavefunctions,
	which, by definition, are analytic and represent
	stationary modes of the waves in eq.\ (\ref{e:CR2Wave}).
	Moreover,
	the wavefunctions are amplitudes of probabilities
	that are equivalent to information in Shannon theory,
	so that
	the stationary modes literally represent physical information
	that would become available on measurement.
	We do not need empirical authority, therefore,
	beyond that which already led to
	the non-relativistic quantum theory,
	in order to identify our $c$ with
	that in quantum field theory,
	and, as a special case, with the speed of light,
	also establishing inherent consistency
	between relativity and quantum mechanics
	(\S\ref{s:scale}).
\NE

General relativity concerns
the complementary case of possibly \emph{comoving} observers
unable to share their referents
because of a spatial or temporal separation $x^\mu$,
which means that
the physical conditions in their neighbourhoods 
can no longer be assumed to be identical,
so that linearity cannot hold.
Analyticity is still a valid assumption,
yielding the general curvilinear transformation
which is the basis of general relativity,
\begin {equation} \label {e:xfmMetric}
	\GR(x^\mu): \quad
		d{x'}^\nu
		=
			g_{\nu}^\mu \, dx^\kappa
		,
	\quad
		d{s'}^2
		=
			g_{\nu\kappa}(x^\mu)
			\;
			d{x'}^\nu
			\,
			d{x'}^\kappa
		.
\end {equation}
The relativity of scale is thus adequate logical foundation
for both theories of relativity.
\Qed


\end {document}

%% file: tbphub.tex
\begin {table}[htbp]
\begin {minipagetbl}{3.25in}
\small
\begin {tabular}{|| l | r@{}r@{}l@{} | r@{}r@{}l@{} | r@{}r@{}l@{} ||}
\hline
\emph{Body}
& \multicolumn{3}{c|} {\emph{Distance}}
& \multicolumn{6}{c||}{\emph{Recession} (m/y)}	\\
& \multicolumn{3}{c|} {(AU)}
& \multicolumn{3}{c}  {$h_c = 50^\dag$}
& \multicolumn{3}{c||}{$h_c = 75^\dag$}				\\
\hline
Moon$^\ddag$
	& $2.57$&$\times$&$10^{-3}$
		& 	0.020&&	&	0.029&&\  \\
\hline
Mercury	&	0.40&&	&	3.0&&		&	4.4&&	\\
Venus	&	0.72&&	&	5.5&&		&	8.3&& 	\\
Earth	&	1.00&&	&	7.6&&		&	11.5&& 	\\
Mars	&	1.52&&	&	11.7&&		&	17.5&& 	\\
\hline
Jupiter	&	5.20&&	&	39.8&&		&	59.7&& 	\\
Saturn	&	9.56&&	&	73.1&&		&	109.6&&	\\
Uranus	&	19.19&&	&	146.8&&		&	220.2&&	\\
Neptune	&	30.11&&	&	230.3&&		&	345.5&&	\\
Pluto	&	39.53&&	&	302.4&&		&	453.5&&	\\
\hline
\end {tabular}
\footnotetext[0]{$^\dag$ in km/s-Mpc,
	$= 1.62, 2.43\order{-18}$~\s{}, respectively. }
\footnotetext[0]{$^\ddag$ from earth;
	measured: $0.0384$~m/y (Dickey \etal, 1994)}
\end {minipagetbl}

\caption {Planetary Hubble flow}
\label {t:planetary}
\end {table}